\documentclass[preprint,letter]{IEEEtran}

\usepackage{fancyhdr}
\usepackage{cite}
\usepackage{mathptmx} 			
\usepackage{mathtools}
\usepackage{amsmath}                	
\usepackage{amsthm}                 	
\usepackage{amssymb}
\usepackage{enumerate}
\usepackage{dsfont}
\usepackage{wasysym}
\usepackage{mathdots}
\usepackage[mathscr]{euscript}
\usepackage{graphicx}
\usepackage{eurosym}                	
\usepackage{lscape}                 		
\usepackage{longtable}
\usepackage{multirow}
\usepackage{subfigure}
\usepackage{rotating}
\usepackage{color,soul}                  	
\usepackage[colorlinks=true, citecolor=magenta]{hyperref}       
\usepackage{enumerate}
\usepackage{booktabs}
\usepackage{array}
\usepackage{comment}
\usepackage{nomencl}

\theoremstyle{definition}

\theoremstyle{remark}

\DeclareMathOperator*{\argmin}{arg\,min}

\title{A Distributed Economic Model Predictive Control Design for a Transactive Energy Market Platform in Lebanon, NH}
\author{Steffi~Olesi~Muhanji,~\IEEEmembership{Student Member,~IEEE,}
         Samuel V. Golding, Tad Montgomery, Clifton Below, and~Amro~M~Farid,~\IEEEmembership{Senior Member,~IEEE}
\thanks{S. O. Muhanji is with the Thayer School of Engineering at Dartmouth, Hanover, NH, 03755 USA email: steffi.o.muhanji.th@dartmouth.edu.}
\thanks{S. Golding is President and Co-Founder of Community Choice Partners Inc., Concord, NH USA email: golding@communitychoicepartners.com.}
\thanks{T. Montgomery is the Energy and Facilities Manager with the City of Lebanon, NH 03766, USA email: Tad.Montgomery@lebanonnh.gov.}
\thanks{C. Below is the Assistant Mayor with the City of Lebanon, NH, 03755 USA email: Clifton.Below@lebanonnh.gov.}
\thanks{A. M. Farid an Associate Professor with the Thayer School of Engineering at Dartmouth, Hanover, NH, 03755 USA email: amro.farid@dartmouth.edu.}
\thanks{Manuscript received October 31, 2020; revised ------------.}}

\begin{document}

\maketitle

\begin{abstract}
The electricity distribution system is fundamentally changing due to the widespread adoption of variable renewable energy resources (VREs), network-enabled digital physical devices, and active consumer engagement. VREs are uncertain and intermittent in nature and pose various technical challenges to power systems control and operations thus limiting their penetration. Engaging the demand-side with control structures that leverage the benefits of integral social and retail market engagement from individual electricity consumers through active community-level coordination serves as a control lever that could support the greater adoption of VREs. This paper presents a Distributed Economic Model Predictive control (DEMPC) algorithm for the electric power distribution system using the augmented lagrangian alternating direction inexact newton (ALADIN) algorithm. Specifically, this DEMPC solves the Alternating Current Optimal Power Flow (ACOPF) problem over a receding time-horizon. In addition, it employs a social welfare maximization of the ACOPF to capture consumer preferences through explicit use of time-varying utility functions. The DEMPC formulation of the ACOPF applied in this work is novel as it addresses the inherent dynamic characteristics of the grid and scales with the explosion of actively controlled devices on the demand-side. The paper demonstrates the simulation methodology on a 13-node Lebanon NH distribution feeder.
\end{abstract}

\section{Introduction}


In recent years, significant attention has shifted towards the effective technical and economic control of the electricity distribution system to address the complex challenge of operating electricity grids with large amounts of variable renewable energy resources (VREs) such as solar and wind. This shift in focus has been driven primarily by the rapid evolution of distribution grids to include: 1) a more active consumer base, 2) numerous smart digital devices, and 3) large amounts of distributed energy resources (DERs)\cite{Muhanji:2018:SPG-J37}. Unfortunately, the ambitious  goal of decarbonizing the electric power grid while enhancing its sustainable and resilient operation presents technical, economic, and regulatory challenges.  

The first of these technical challenges is that the uncertain and intermittent nature of VREs appears over multiple timescales and horizons\cite{Farid:2014:SPG-J26}. This necessitates control techniques that capture the inter-timescale dynamics introduced to the electricity net load by VREs\cite{Muhanji:2018:SPG-J37}. In that regard, numerous model predictive control (MPC) algorithms -- centralized as well as distributed -- have been proposed for power systems applications within the context of VRE integration. MPC is an optimal feedback control technique that uses the dynamic state of a system to predict over a finite and receding time horizon how the state of the system evolves and uses only the solution for the first time-step to update the system for the next optimization block\cite{Ellis:2014:00}. This feedback-based closed-loop control helps to compensate for the net-load variations and stochasticity introduced by VREs in real-time operations\cite{Xie:2009:01}. A majority of the proposed centralized applications have focused on the dynamic economic dispatch problem \cite{Xia:2009:00,Xia:2011:00,Xie:2009:01} for systems with a high penetration of VREs or on optimal dispatch of DERs for distribution system microgrids\cite{Zhu:2014:00, Mayhorn:2012:00}. In the meantime, decentralized approaches explore similar themes as centralized ones with most focusing on the economic dispatch problem\cite{Velasquez:2019:00} or environmental dispatch with intermittent generation resources\cite{Xie:2009:01, Alejandro:2014:00}. However, a recent study has shown that the convergence to optimal values is not always guaranteed for decentralized approaches and that a majority of these studies neither consider ramping rates nor the impact of VREs on dispatch decisions\cite{Velasquez:2019:00}.

The second of these technical challenges is that a high penetration of VREs undermines the dispatchability of the generation fleet and, therefore, requires the activation of demand side resources. Traditionally, the generation fleet comprised of large controllable thermal power plants meant to serve fairly passive loads. However, as more and more VREs are added to the electricity grid, the variability of the system net load increases significantly introducing with it dynamics that span multiple timescales. The term ``net load'' here is defined as the forecasted demand minus the forecasted variable generation from wind and solar. This means that in real-time operations, controllable generators must not only compensate for net load forecast errors but also provide extensive ramping capability to account for changes in variable generation due to external factors such as solar irradiance and wind speed. In the meantime, there are fewer dispatchable generators to serve this balancing role.  This two-fold technical challenge greatly limits the penetration of VREs and, therefore, calls for more highly responsive control levers. Activating the demand-side is seen as the remaining potential control lever given its evolution to include: 1) an active consumer base, 2) numerous smart energy internet of things (eIoT) devices, and 3) large amounts of distributed energy resources (DERs). These three factors increase the controllability of the demand-side paving the way for various demand-side management (DSM) solutions that can be used to shift, shed, and/or increase electricity demand in the real-time in order to balance variations in net load.


The dynamic nature of VREs also necessitates frequent decision-making which requires automated (rather than manual) solutions on distributed edge devices called the energy Internet of Things (eIoT).  This frequent decision-making requires robust information and communication technologies (ICTs) that enable intelligent coordination of these distributed eIoT devices\cite{Vrba:2014:00}. eIoT solutions must scale with the number of devices, deal with computational complexity and handle communication with other distributed devices in a timely fashion\cite{Vrba:2014:00, Farid:2015:SPG-J17}.  Multi-agent systems (MAS) have been proposed in the literature to address the practical challenges of controlling a large number of active grid edge devices in the short time span of power grid markets\cite{Vrba:2014:00} . Smart devices whether it is rooftop solar, electric vehicles (EVs), programmable thermostats, or battery energy storage, can coordinate as agents within a MAS to reach a global consensus that maintains power system balance or stability. In MAS approaches, agents can simplify decision making by communicating with only their neighbours to make local decisions that inform higher-level decisions\cite{Rivera:2014:SPG-BC01, Toro:2015:00, Santos:2015:01}. This significantly reduces the amount of shared information among agents and also allows for a more robust system by eliminating the single point of failure. At the core of MAS applications are distributed control algorithms that are employed to solve local sub-problems so as to reach consensus on global objectives. 

The integration of demand side resources at the grid periphery begets a third challenge; the shear number.  The demand-side is comprised of millions or even billions of actively interacting cyber-physical devices that are distributed both spatially as well as functionally\cite{Vrba:2014:00, Siano:2016:00}. Controlling these devices requires correspondingly distributed and scalable control algorithms\cite{Farid:2015:SPG-J17}. Distributed control algorithms have been proposed as solutions that can scale up to such a large number of devices and still be implemented in the minute-timescale of power system markets\cite{Muhanji:2018:SPG-J37}. Through effective coordination, distributed control algorithms can be used to coordinate local sub-problems to reach a global objective similar to that achieved by centralized algorithms\cite{Muhanji:2018:SPG-J37}. 


In addition, these algorithms must respect the physical constraints of the grid which are both non-linear and non-convex. The optimal power flow (OPF) problem is among the most common optimization problems used in the economic control of the power system\cite{Exposito:2016:00}. The OPF determines the optimal flows of power through a given electricity network to meet demand and respect operational constraints. 
Several variants of the OPF problem exist\cite{Qiu:2009:00, Frank:2012:01, Frank:2012:01}; the alternating current (AC) OPF variant uses the full implementation of the ``power flow equations" which, in turn, are a pseudo-steady state model of Kirchkoff's current law\cite{Exposito:2016:00, Frank:2012:00, Qiu:2009:00} and is thus, non-linear and non-convex.
%
As one would expect, various distributed control algorithms have also been proposed for the OPF problem\cite{Molzahn:2017:01}. 
However, due to the non-linear, non-convex nature of the ACOPF, a majority of these algorithms seek to either linearize the ACOPF or use other relaxation techniques such as semi-definite programming (SDP)\cite{Molzahn:2016:00, Molzahn:2013:00} or second-order cone programming (SOCP). While such mathematical simplifications have their algorithmic merits, they often fail to fully capture the complex and dynamic behaviour of distribution systems\cite{Molzahn:2017:01}. Additionally, many of the proposed algorithms such as the Alternating Direction Method of Multipliers (ADMM), Alternating Target Cascading (ATC), and Dual Ascent have practical implementation weaknesses that make them unreliable when applied to large-scale applications\cite{Molzahn:2017:01}. The most common of these algorithms is the ADMM which has been widely studied in the literature in its application to the electric power grid\cite{Erseghe:2014:00, Guo:2017:00}. Unfortunately, recent studies have shown that the convergence of the ADMM depends highly on the choice of tuning parameters in convex spaces and is all-together not guaranteed in non-convex spaces such as the ACOPF\cite{Molzahn:2017:01}. In recent years, the Augmented Lagrangian Alternating Direction Inexact Newton (ALADIN) algorithm has been proposed in the literature as not just an alternative to the ADMM but also as a solution with better convergence guarantees even for non-convex applications such as the ACOPF\cite{Houska:2016:00, Houska:2017:00}.


To be successful on a practical level, in addition to the technical challenges above, the distributed control algorithm must be implemented within an appropriate commercial  and regulatory framework. Community choice aggregation (CCA) represents one such framework, and is authorized in California, Massachusetts, New York, New Jersey, Illinois, Ohio, Rhode Island and New Hampshire\cite{OShaughnessy:2019:00}. It is a policy that allows local governments (e.g. towns, cities and counties) to become the default electricity provider and enroll customers within their municipal boundaries that are currently on utility basic service on an opt-out basis\cite{OShaughnessy:2019:00}.
CCAs compete on the basis of electricity procurement and retail innovation by offering consumers access to a broader portfolio of electric products, often at more competitive prices than those traditionally offered by utilities \cite{Kuo:2014:00, OShaughnessy:2019:00}. CCAs are thus naturally incentivized to facilitate retail demand flexibility and the intelligent management of distributed energy to create revenue streams in new ways, by integrating these assets into wholesale market operations, the CCA’s portfolio risk management, and distribution company network planning and operations. CCAs are, therefore, also incentivized to advocate for the regulatory reforms necessary to value and monetize Distributed Energy Resources in ways that account for their temporal and geographic attributes, and to expand data interchange and market access for innovative third-party companies. CCAs in certain states, most notably in California, have consequently focused on expanding retail programs and third-party customer services, and engaged in multi-sectoral decarbonization planning and local infrastructure development (e.g. microgrids, non-wires alternatives)\cite{Local-Self-Reliance:2020:00}. However, CCAs may face operational barriers to retail innovation due to the statutory requirement that distribution utilities continue to provide retail meter reading, data management and consolidated billing functions\cite{Utilities:2020:00}. The New Hampshire market is distinguished as the only state wherein the statutory authorities of CCAs allow for the direct provision of the aforementioned retail customer services, which are critical to enabling Transactive Energy. Consequently, CCAs in New Hampshire represent a viable commercial pathway to overcome legacy utility IT systems and implement the concept of a shared integrated grid that is characterized by: 1) integral social and retail market engagement from electricity consumers, 2) the digitization of energy resources with the eIoT, and 3) widespread community-level coordination\cite{Muhanji:2019:EWN-C67}. Towards this end, the City of Lebanon and other interested municipalities are drafting a Joint Power Agreement\cite{Lebanon-NH:2020:00} to create an agency called ``Community Power New Hampshire''\cite{CPNH:2020:00} that will offer operational services to all CCAs on a statewide basis, and have already begun engaging in regulatory proceedings to create the market and control structures necessary to enable the efﬁcient and low-cost exchange of energy data, products and services\footnote{Refer to filings submitted by the City of Lebanon and the Local Government Coalition in NH PUC Docket 19-197. Available online: https://www.puc.nh.gov/Regulatory/Docketbk/2019/19-197.html}.

To increase consumer participation, CCAs must provide grid services that engage consumers and allow for the expression of their preferences. Typically, the bulk of consumers at the distribution system are residential homes. These consumers generally represent small loads and are driven by factors such as comfort, ease of use, and cost. This naturally demands market and control structures within CCAs that ultimately enable the efficient, and low-cost exchange of electricity products and services among consumers. These market and control structures must recognize that the value of electricity demand changes not just with quantity but also with the time of day.  For instance, a commercial supermarket may be unwilling to shed 1kw of consumption for refrigeration at 7am as they are opening but could shed 1kw for laptop computers in the middle of the day after their batteries have been charged.  Similarly, someone with a set routine may be willing to pay more for a hot-water shower in the morning than for the same shower in the afternoon. Given the time and usage value of electricity, transactive energy market models implemented by CCAs must capture the social benefits to consumers by explicitly implementing time-varying utility functions.

\vspace{-0.1in}
\subsection{Contribution}
Given these many technical, economic, and regulatory considerations, this paper develops a distributed transactive energy control system for the economic control of an electric power distribution system. It offers several key novel features relative to the existing literature. (1) Unlike the traditional single time step ACOPF problem based upon algebraic constraints, this work recasts the ACOPF formulation into an economic MPC with a finite look-ahead time horizon and explicit state variables.  Consequently, the system proactively responds to the variability of the net load while controlling the energy stored within the distribution system. (2) The objective function in this work minimizes social welfare and incentivizes demand-side participants to have elastic behavior. (3) Demand-side utility functions applied in this study are also explicitly time-varying to account for consumer's preferences changing over the course of the day. (4) To account for the potential explosion of active devices at the grid's periphery, the EMPC problem is implemented as a multi-agent control system based on the ALADIN algorithm which has been proven to converge to a local minimizer even for nonlinear, non-convex constraints such as those presented by the ACOPF equations.  (6) Finally, the DEMPC is tested on a 13-bus feeder from the City of Lebanon, NH in which controllable demand, controllable generation, stochastic generation and stochastic demand resources have been added.
\vspace{-0.1in}
\subsection{Outline}
The rest of this paper is organized as follows: In Section \ref{sec:acopf},  the ACOPF problem, in its generic form, is presented.  Section \ref{sec:empc} introduces a generic formulation of economic MPC problem.  Section \ref{sec:acopfmpc} outlines the ACOPF problem reformulated as an economic MPC with a social welfare minimization to capture consumer preferences.  Section \ref{sec:aladin} then introduces the ALADIN algorithm and discusses its application to the previously mentioned EMPC ACOPF model. Section \ref{sec:simres} numerically demonstrates the convergence of ALADIN to the EMPC-ACOPF model for a 13-bus feeder for the City of Lebanon, NH and provides a discussion of the results. Finally, the paper is concluded in Section \ref{sec:conc}.  
\vspace{-0.1in}
\section{Background}\label{sec:background}
\subsection{The AC Optimal Power Flow Problem}\label{sec:acopf}
The ACOPF calculates the steady-state flows of power within any given electrical network. It is comprised of an objective function typically a generation cost minimization and is constrained by generation capacity limits, voltage magnitude limits, and power flow constraints. The traditional ACOPF formulation is presented below:
\vspace{-0.1in}
\begin{align}
    min \quad &C(P_{GC}) = P_{GC}^TC_2P_{GC} + C_1^TP_{GC} +C_0\mathbf{1}\label{eq1:costfunc}\\
    s.t. \quad &A_{GC}P_{GC} -A_{DS}\hat{P}_{DS} = Re\{diag(V)Y^*V^*\}\label{eq2:pbP}\\
    &A_{GC}Q_{GC} - A_{DS}\hat{Q}_{DS} = Im\{diag(V)Y^*V^*\}\label{eq3:pbQ}\\
    &P_{GC}^{min}\leq P_{GC}\leq P_{GC}^{max}\label{eq4:Plims}\\
    &Q_{GC}^{min}\leq Q_{GC}\leq Q_{GC}^{max}\label{eq5:Qlims} \\
    &V^{min}\leq |V|\leq V^{max}\quad \label{eq6:Vlims}\\
    &e_x^T\angle V= 0&\label{eq7:refa}
\end{align}
\vspace{-0.1in}
The following notations are used in this formulation:
\begin{align*}
&GC                                 &&\text{index for controllable generators}\\
&DS                                 &&\text{index for traditional demand units}\\
&C_{2}, C_{1}, C_{0}                &&\text{quadratic, linear, and fixed cost terms}\\
&                                   &&\text{of the generation fleet}\\
& e_x                               && \text{reference angle elementary basis vector}   \\
& P_{GC}, Q_{GC} 	                && \text{active/reactive power generation}\\
& \hat{P}_{DS},	\hat{Q}_{DS}	    && \text{total forecasted active/reactive demand}\\
& P_{GC}^{min},  P_{GC}^{max}	    && \text{min/max active generation limits}\\
& Q_{GC}^{min}, Q_{GC}^{max}	    && \text{min/max reactive generation limits}\\ 
& V^{min}, V^{max}	                && \text{min/max voltage limits at buses}
\end{align*}
\begin{align*}
& Y                                 && \text{bus admittance matrix}\\
& V                                 && \text{bus voltages}\\
& N_G				                && \text{number of generators}\\
& A_{GC}                            && \text{generator-to-bus incidence matrix}\\
& A_{DS}                            && \text{load-to-bus incidence matrix}
\end{align*}
 Equation \ref{eq1:costfunc} represents the quadratic generation cost function.  Note that $C_{2}$ is a diagonal matrix and so the generation cost objective function is separable by generator. This cost function may be equivalently written as:  
\begin{equation}
    C(P_G)=\sum_{g\in G} c_{2g}P_g^2 + c_{1g}P_g + c_{0g}\label{eq:cf}
\end{equation}
Equations \ref{eq2:pbP} and \ref{eq3:pbQ} define the active and reactive power flow constraints at a bus respectively. While Equations \ref{eq4:Plims}, \ref{eq5:Qlims}, and \ref{eq6:Vlims} represent the active power generation, reactive power generation and bus voltage limits.  Finally, \ref{eq7:refa} sets the voltage angle of the chosen reference bus(es) to 0.

\subsection{A Generic Non-linear Economic MPC Formulation} \label{sec:empc}
MPC is an optimization-based control algorithm that solves a dynamic optimization problem over a receding time horizon of T discrete time steps. The solution to the optimization problem is computed over k=[0,\ldots, T-1] and the solution for $k=0$ is applied to the control input u[k=0].  The clock is then incremented and the same process is repeated over k=[1,\ldots, T] and so on. 
A generic non-linear economic model predictive control algorithm is presented below\cite{Ellis:2014:00}.

\vspace{-0.2in}
\begin{align}
\argmin_{u_{k=0}} \quad & J = \sum_{k=0}^{T-1} x_k^TQx_k + u_k^TWu_k^T + Ax_k + Bu_k&\label{eq:costfuncmpc} \\
s.t.\qquad & x_{k+1}=f(x_k,u_k,\hat{d}_k)&\label{eq:state}\\  
 & u_{min}\leq u_k \leq u_{max}&\label{eq:cap1}\\ 
 & x_{min}\leq x_k \leq x_{max}&\label{eq:cap2}\\ 
 & x_{k=0}= \tilde{x}_{0}&\label{eq:initc}
\end{align}
\vspace{-0.2in}
\begin{align*}
& \hat{d}_k 	                    && \text{predicted disturbance at discrete time $k$} \\
& x_k, u_k                          && \text{system states and inputs at time $k$}   \\
& x^{min}, x^{max}	                && \text{min/max system state limits}\\
& u^{min}, u^{max}	                && \text{min/max system input limits}
\end{align*}
whereby Equation \ref{eq:costfuncmpc} represents the economic objective function, Equation \ref{eq:state} defines the non-linear dynamic system state equation while Equations \ref{eq:cap1}, and \ref{eq:cap2} define the capacity constraints for the system inputs and states respectively. Lastly, Equation \ref{eq:initc} defines the initial conditions.  

\section{Methodology and Simulation Setup}
\subsection{An Economic MPC Formulation of a Multi-Period  AC Optimal Power Flow}\label{sec:acopfmpc}
The ACOPF formulation in Section~\ref{sec:acopf} lacks several features: 1.) a multi-time period formulation, 2.) ramping constraints on generation units, 3.) controllable demand and stochastic generation units , 4) a time-varying demand-side utility function, and 5) an explicit description of system state.  The last of these requires the most significant attention. The power flow equations in Equations \ref{eq:pbPm} and \ref{eq:pbQm} are derived assuming the absence of power grid imbalances and energy storage\cite{Exposito:2016:00}.  In reality, however, all power system buses are able to store energy; even if it be in relatively small quantities.  Consequently, relaxing the inherent assumptions found in the traditional power flow equations introduces a state variable $x_k$ associated with the energy stored at the power system buses during the $k^{th}$ time block.  Naturally, limits are imposed on this state variable to reflect the physical reality and an initial state $\tilde{x}_0$ is included in the EMPC ACOPF formulation.
\begin{align}
\argmin_{P_{GCk=0}} \;\; &J = \sum_{k=0}^{T-1 } \big[C_{GC}(P_{GCk}) + C_{DCk}(P_{VGk}))\big]&\label{eq:costf}\\
&P_{VGk} = \hat{P}_{DCk} -P_{DCk}&\label{eq:pvpp}\\
&Q_{VGk} = \hat{Q}_{DCk} -Q_{DCk}&\label{eq:qvpp}\\\nonumber
s.t. \quad &x_{k+1}=x_k+ \Delta T \big(A_{GC}P_{GCk} +  &\\\nonumber
&\quad \ldots A_{GS}\hat{P}_{GSk} +  A_{DC}P_{VGk}-A_{DS}\hat{P}_{DSk}- &\\
&\quad \quad \quad \quad \quad \ldots Re\{diag(V_k)Y^*V_k^*\}\big)&\label{eq:pbPm}\\\nonumber
&0=A_{GC}Q_{GCk} +  A_{DC}Q_{VGk}- &\\
&\quad \quad \ldots A_{DS}\hat{Q}_{DSk} - Im\{diag(V)Y^*V^*\}&\label{eq:pbQm}\\
&P_{GC}^{min}\leq P_{GCk}\leq P_{GC}^{max}&\label{eq:plimits}\\
&P_{VGk}^{min}\leq P_{VGk}\leq P_{VG}^{max}&\label{eq:pdlimits}\\
&Q_{VGk}^{min}\leq Q_{VGk}\leq Q_{VG}^{max}&\label{eq:qdlimits}\\
&Q_{GC}^{min}\leq Q_{GCk}\leq Q_{GC}^{max}&\label{eq:qlimits}\\
& \Delta T R_{GC}^{min}\leq P_{GCk}-P_{GC,k-1}\leq \Delta T R_{GC}^{max}&\label{eq:rlimits}\\
&V^{min}\leq |V_k|\leq V^{max}&\label{eq:vlimits}\\
&x^{min}\leq x_k\leq x^{max}&\label{eq:xlimits}\\
&e_x^T\angle V_k= 0&\label{eq:refb}\\
& x_{k=0}= \tilde{x}_{0}&\label{eq:initcond}
\end{align}
\vspace{-0.3in}
\begin{align*}
&GS                             &&\text{stochastic generators index}\\
&DC                             &&\text{index for controllable demand units}\\
&\mathbb{A}_{DCk}, \mathbb{B}_{DCk}, \mathbb{C}_{DCk}  &&\text{quadratic, linear, fixed cost terms of}\\
&                               &&\text{controllable demand at time k.}\\
& R_{GC}^{min},R_{GC}^{max}	    && \text{max/min generation ramp limits}\\
& P_{GCk}, Q_{GCk}	            && \text{active/reactive controllable generation at $k$ }\\
& P_{GSk} 	                    && \text{active stochastic generation at time $k$ }\\
& \hat{P}_{DSk}, \hat{Q}_{DSk}	&& \text{active/reactive demand forecast at time $k$}\\
& \hat{P}_{DCk}, \hat{Q}_{DCk}	&& \text{active/reactive forecasted controllable}\\
&                               &&\text{demand at discrete time $k$}\\
& {P}_{DCk}, {Q}_{DCk}	        && \text{active/reactive dispatched controllable}\\
&                               &&\text{demand at discrete time $k$}\\
& x_{k+1}                       &&\text{system state at time $k+1$}\\
&\Delta T                       &&\text{time step of the optimization}\\
& A_{GS}, A_{DC}                && \text{stochastic generator-to-bus \& controllable }\\
&                               && \text{load-to-bus incidence matrices }
\end{align*}
Note that this EMPC ACOPF formulation is equivalent to the traditional ACOPF when $T=0$, $x^{min}= x^{max}=0$ and $R_G^{min},R_G^{max} \rightarrow \infty$. The objective function has also been modified to minimize the overall cost of controllable generation and the cost of virtual generation for T discrete time-steps. Notice that the cost of controllable generation remains the same as before and is given by:
\begin{align*}
\vspace{-0.1in}
  C_{GC} = P_{GCk}^TC_2P_{GCk} + C_1^TP_{GCk}  +C_0\mathbf{1}  
\end{align*}

Similarly, the cost of virtual generation follows a quadratic form as that of controllable generation and defined as follows:
\begin{align*}
  C_{DC} = (\hat{P}_{DCk} - P_{DCk})^T\mathbb{A}_{DCk}(\hat{P}_{DCk} - P_{DCk})+ \ldots\\ 
  \quad \mathbb{B}_{DCk}^T(\hat{P}_{DCk} - P_{DCk})  +\mathbb{C}_{DCk}\mathbf{1}  
\end{align*}
Whereby the coefficients $\mathbb{A}_{DCk}$, $\mathbb{B}_{DCk}$, and $\mathbb{C}_{DCk}$ vary in time to reflect consumer preferences at various points during the day. In addition to these changes, two new energy resources are introduced namely, controllable demand ($\hat{P}_{DCk} - P_{DCk}$) and stochastic generation $\hat{P}_{GSk}$. The virtual generation ($\hat{P}_{DCk} - P_{DCk}$) is also subject to capacity limits given by Equation~\ref{eq:pdlimits}. To eliminate baseline errors associated with virtual power plants\cite{Jiang:2015:SPG-J21}, the capacity limits of virtual generation are set as 20\% of forecasted stochastic demand for each demand node.
\subsection{The ALADIN (Augmented Lagrangian Alternating Direction Inexact Newton) Algorithm}\label{sec:aladin}
\vspace{-0.1in}
\begin{figure}[!h]
\vspace{-0.1in}
    \centering
    \includegraphics[scale=0.4]{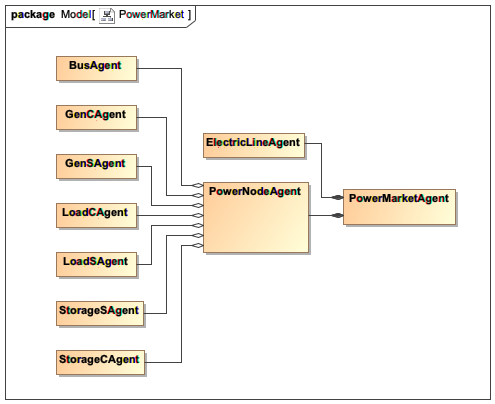}
    \caption{Area agent architecture.}
    \label{fig:areaAgent}
    \vspace{-0.1in}
\end{figure}
The ALADIN algorithm admits an optimization problem of the form: 
\begin{align}
\vspace{-0.1in}
\argmin_{y_i} \quad & J=\sum_{i} f(y_{i})\label{eq:costfunc}\\
s.t. \quad & h_i(y_{ik}) = 0\label{eq:equalityconst}\\ 
&A_iy_{ik} = 0 \label{eq:consensus}\\ 
&y_{i}^{min}\leq y_{i}\leq y_i^{max}&\label{eq:limits}
\vspace{-0.1in}
\end{align}
where the generic objective function $J$ is separable with respect to N sets of decision variables $y_i$.  Furthermore, there is a non-linear, not necessarily convex, function $h_i()$ for each $y_i$.  Equation \ref{eq:consensus} is a linear consensus constraint which serves as the only coupling between the subsets of decision variables.  Finally, Equation \ref{eq:limits} adds minimum and maximum capacity constraints on the decision variables.  The distributed control algorithm for solving the above optimization problem is discussed in full in  \cite{Houska:2017:00} and proven to converge even for cases where the functions $h_i$ are non-linear and/or non-convex.  
\begin{figure}[!h]
    \centering
    \includegraphics[scale=0.4]{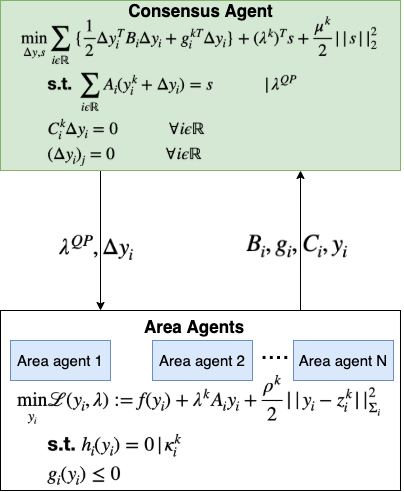}
    \caption{ALADIN agent architecture.}
    \label{fig:aladin}
    \vspace{-0.1in}
\end{figure}
\begin{figure*}[!htbp]
\vspace{-0.1in}
    \centering
    \includegraphics[width=0.85\textwidth]{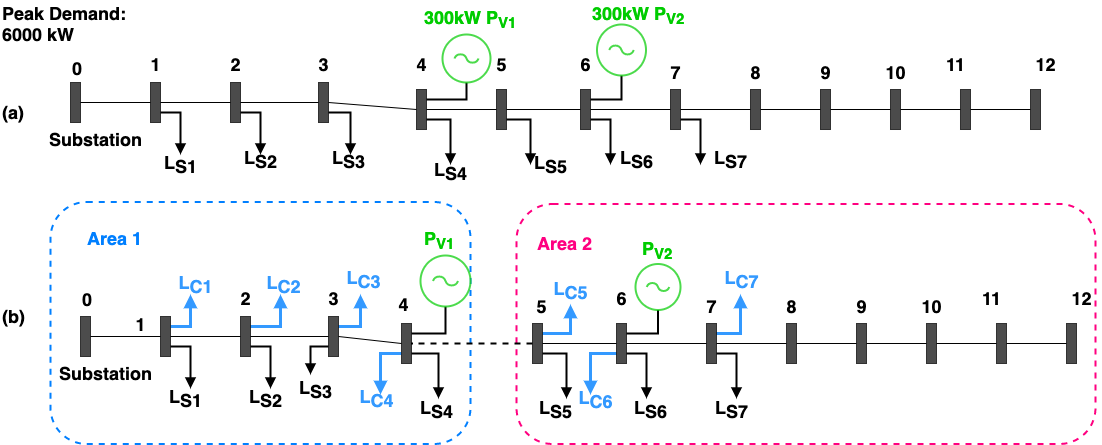}
    \caption{13-bus distribution feeder for the City of Lebanon NH, with a peak demand of 6000kW, 2 300kW solar PV plants and 7 conventional loads.}
    \label{fig:16L4}
    \vspace{-0.1in}
\end{figure*}

The distributed ALADIN algorithm is best summarized by Figure~\ref{fig:aladin}. The algorithm is comprised of two steps, a fully distributed step where area agents compute the solution to a non-linear optimization sub-problem for their respective control area. Each control area represents a power system area with a local agent architecture as the one depicted in Figure~\ref{fig:areaAgent}.  The sub-problem in a given control area is obtained by rearranging Equations \ref{eq:costfunc}, \ref{eq:equalityconst}, \ref{eq:consensus}, and \ref{eq:limits} as shown in Figure~\ref{fig:aladin}. 

The area agents then share their hessians, jacobians, gradients, and local solutions with the consensus agent who then determines the updates ($\Delta y_i$ and $\lambda_{QP}$) for the dual and primal variables  by solving the quadratically-constrained problem (QCP) shown in Figure~\ref{fig:aladin}. Notice that the role of the consensus agent may be carried out by a centralized facilitator or by any of the local area agents. The dual and primal variables are updated according to equations \ref{eq:primalupdate} and \ref{eq:dualupdate}. In some cases, a line search is carried out to determine the update rate for coefficients $\alpha_1, \alpha_2, \text{and }\alpha_3$  otherwise,  $\alpha_1=\alpha_2=\alpha_3=1$.  
\begin{align}\label{eq:primalupdate}
\vspace{-0.1in}
&z^{k+1} \leftarrow z^k+\alpha_1^k(y^k-z^k)+\alpha_2^k\Delta y^k&\\\label{eq:dualupdate}
&\lambda^{k+1} \leftarrow \lambda^k+\alpha_3^k(\lambda^{QP}-\lambda^k)&
\vspace{-0.1in}
\end{align}
Two penalty parameters $\rho$ and $\mu$ are employed in this algorithm for the local sub-problems and the consensus QCP respectively. These parameters are updated according to Equation \ref{eq:penaltyupdate}. $r_{\rho}$ and $r_\mu$ are constants that are chosen specifically to aid in updating the penalty parameters.
\begin{align}\label{eq:penaltyupdate}
\vspace{-0.1in}
&\rho^{k+1} (\mu^{k+1})  = 
     \begin{cases}
      r_\rho \rho^k \quad (r_\mu \mu^k) &\text{if } \rho^k < \bar{p}\text{ }(\mu^k < \bar{\mu})\\
      \rho^k (\mu^k) &\text{otherwise} \\
     \end{cases}&
     \vspace{-0.1in}
\end{align}
The EMPC ACOPF problem presented in Section~\ref{sec:acopfmpc} is now solved using the ALADIN algorithm as a distributed control approach.   In order to do so, the decision variables $y = [P_{Gk}; Q_{Gk}; |V_k|; \angle V_k] \quad \forall k=[0,\ldots,T-1]$ are partitioned into several sets of decision variables $y_i=[P_{Gi}; Q_{Gi}; |V_i|; \angle V_i]\quad  \forall i=[1,\ldots, N]$; each corresponding to a predefined control area $i$.  The objective function in Equation \ref{eq:costf} is then recast in a separable form as in Equation \ref{eq:cf} with each generator assigned to a specific control area.  The state equations in Equations \ref{eq:pbPm} and \ref{eq:pbQm} are further partitioned by control area and constitute the non-linear, non-convex functions $h_i()$.  At this point, the consensus constraints in Equation \ref{eq:consensus} serve to ensure that the power flowing from one control area $i_1$ to another control area $i_2$ is equal and opposite to the power flowing from $i_2$ to $i_1$.  The remaining constraints of the EMPC ACOPF problem map straightforwardly to the capacity constraints of the ALADIN optimization problem.  \cite{Engelmann:2018:00, Engelmann:2018:01, Engelmann:2017:00} provide further background explanation of how the ALADIN optimization problem maps to a traditional ACOPF formulation and \cite{Houska:2016:00} discusses the general ALADIN algorithm including a line search implementation.  
\vspace{-0.1in}
\begin{figure*}[!h]
\vspace{-0.1in}
    \centering
    \includegraphics[width=0.85\textwidth]{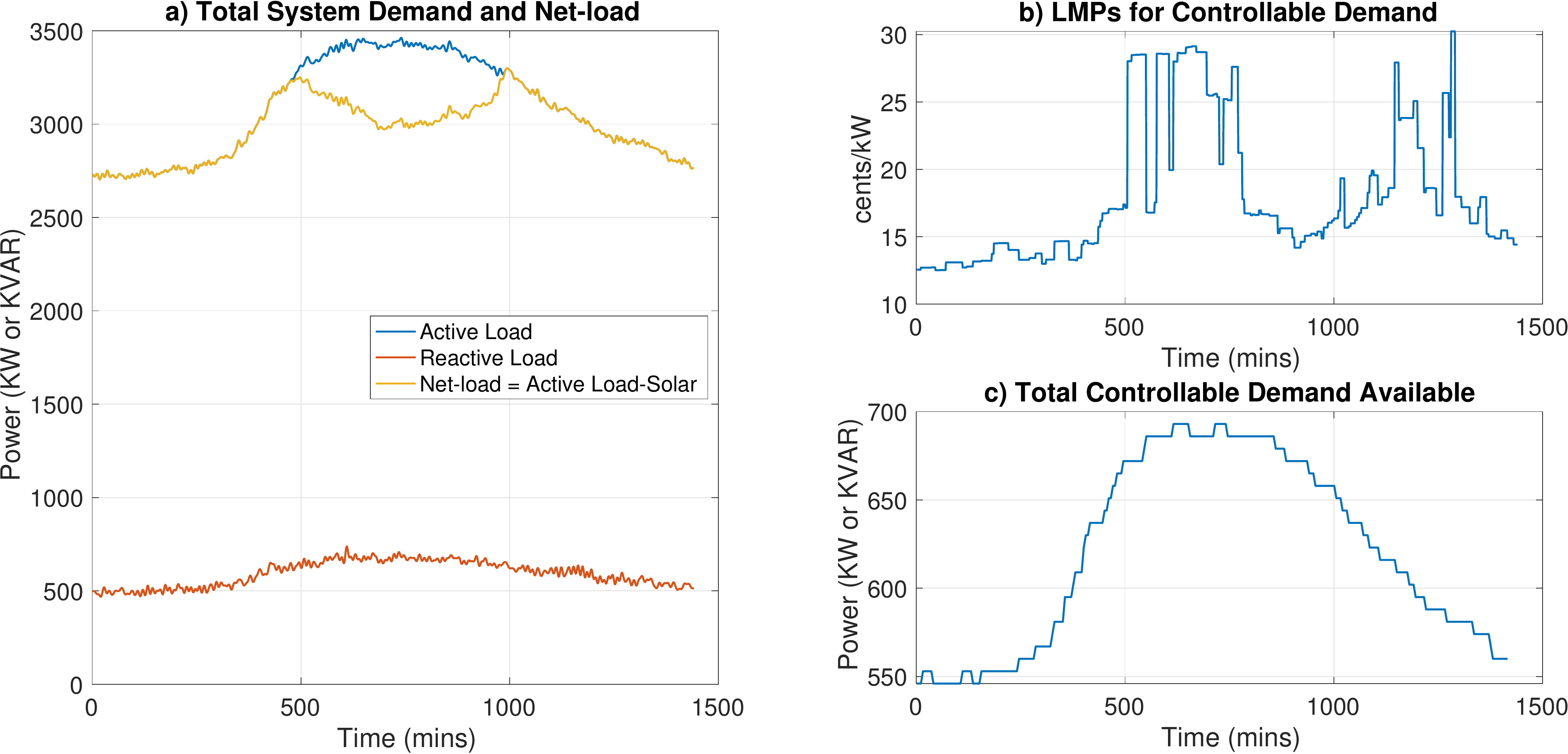}
    \caption{Demand and net load profiles, consensus at boundary buses and convergence of the objective cost value to that of the centralized solution.}
    \label{fig:demProf}
    \vspace{-0.1in}
\end{figure*}
\section{Numerical Demonstration of Convergence}\label{sec:simres} 
The goal of this section is to demonstrate the distributed economic model predictive control design as a potential transactive energy market platform for the City of Lebanon, NH.  More specifically, the DEMPC is numerically demonstrated on real-life data from a 13-bus feeder for the City of Lebanon distribution grid shown in Figure~\ref{fig:16L4}.  (Given the sensitivity of the topology and load data from the local utility, it has not been shared in this publication.)  Figure~\ref{fig:16L4}(a) represents the original feeder with 7 conventional loads [$L_{S1}\rightarrow L_{S7}$] that account for an annual peak load of 6000kW. For the purposes of this study, two solar photo-voltaic (PV) plants each with a capacity of 300kW are placed on nodes 4 and 6. For a distributed simulation, the 13-bus feeder is broken down into two areas as shown in Figure~\ref{fig:16L4}(b). Area 1 is comprised of Nodes 0 to 4 while Area 2 is comprised of Nodes 5 through 12.   To incentivize demand-side participants, virtual power plants [$L_{C1}\rightarrow L_{C7}$] whose maximum capacity is 20\% of the total stochastic demand at the node are added. These plants represents the amount of available controllable demand at each consumer node in time. Note that the maximum capacity limit of the virtual power plants [$L_{C1}\ldots L_{C7}$] changes with time and follows the stochastic demand profile at the individual node. To reach a consensus, the boundary nodes between nodes 4 and 5 must reach the same values for active and reactive power flows as well as angles and voltages for all time steps of the MPC. Additionally, the value of the objective must converge to that of the centralized solution within some error margin. Finally, to test the methodology, an MPC simulation is run every 5-minutes with a 25-min horizon and 5-min time step. Results are presented for a single day.
\begin{figure*}[!htb]
\vspace{-0.1in}
    \centering
    \includegraphics[width=0.85\textwidth]{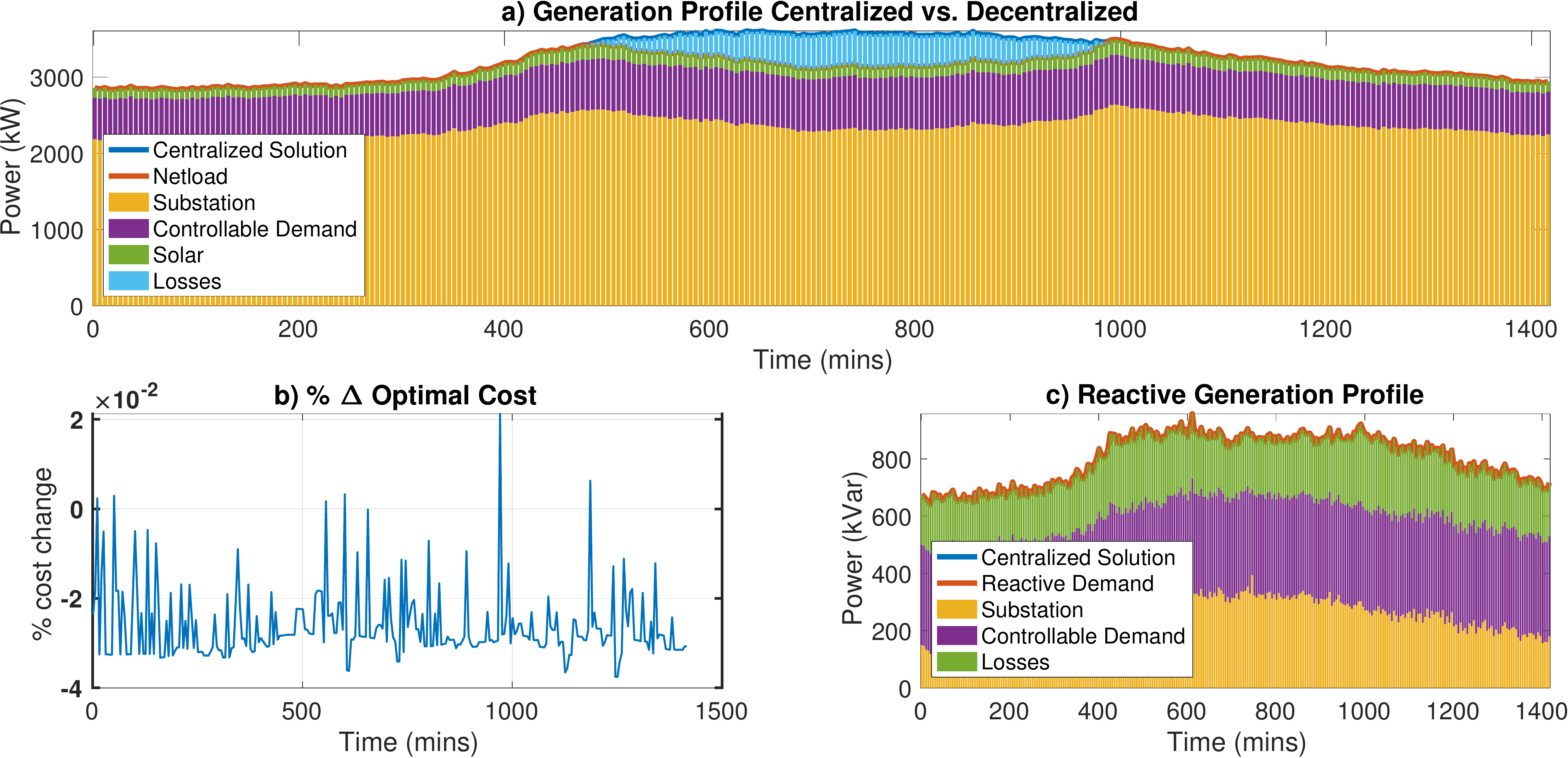}
    \caption{Generation profile comparing the centralized vs. the distributed case and the overall change in optimal cost}
    \label{fig:genProf}
    \vspace{-0.1in}
\end{figure*}
The parameter values used for this study are based on those presented in \cite{Engelmann:2018:00}  and are tweaked as needed. In this study, the two ALADIN penalty parameters $\rho$ and $\mu$ are as follows: $\rho = [1e2 \rightarrow 1e5]$ and $\mu = [1e3\rightarrow 1e5]$. $\rho$ is incremented by a factor of $1.5$ after each iteration while $\mu$ is incremented by a factor of $2$. A line search was not implemented for this demonstration, however, for more complex applications, a line search is recommended to determine the dual and primal update steps\cite{Houska:2016:00}.  The active and reactive demand and net load profiles used in this study are shown in Figure\ref{fig:demProf}(a). The time-varying locational marginal prices (LMPs) that are applied for the virtual power plants are shown in Figure\ref{fig:demProf}(c). Finally, Figure\ref{fig:demProf}(c) represents the total controllable demand available in the system. 


Figure~\ref{fig:genProf}(a) compares the active generation profile from the ALADIN EMPC implemention to that of the centralized EMPC approach. As seen in Figure~\ref{fig:genProf}(a) the ALADIN solution meets demand and results in a final generation profile that matches that of the centralized solution. The active power losses account for approximately 4-6\% of the total demand on the feeder. This result is typical for distribution systems. A comparison of the optimal cost for the centralized versus the distributed approach (illustrated by Figure~\ref{fig:genProf}(b) )shows similar values with a maximum deviation of 0.0212\% from the centralized solution. These results indicate that the solution of the distributed approach closely matches that of the centralized approach with small variations that can be resolved with better parameter estimation and a line search. Finally, Figure~\ref{fig:genProf}(c) shows the reactive power generation profile. Similarly, this figure illustrates that the reactive power demand on the system is met and that the centralized and distributed solutions closely match. 
\vspace{-0.2in}
\section{Conclusion}\label{sec:conc}
 This paper has presented the mathematical formulation for the ACOPF as an EMPC in the context of managing distribution electricity grids with high penetrations of VREs as well as controllable demand. Inherent to the formulation is an introduction of a non-zero energy storage quantity at each bus as a state variable with capacity constraints. This EMPC ACOPF formulation is then recast as a distributed control problem for which the ALADIN algorithm is applied. The paper then demonstrates the methodology on a 13-bus feeder for the City of Lebanon, NH comprising of four types of energy resources, controllable demand and generation, and stochastic demand and generation. The distributed solution is shown to converge to a solution that meets demands and matches the centralized solution. Finally, optimal cost results of the distributed approach closely match those of the centralized solution within a small margin of error.
 \vspace{-0.2in}
\bibliographystyle{IEEEtran}
\bibliography{thesis_work,LIINESLibrary,LIINESPublications}
\vspace{-0.6in}
\begin{IEEEbiography}[{\includegraphics[width=1in,height=1.25in,clip,keepaspectratio]{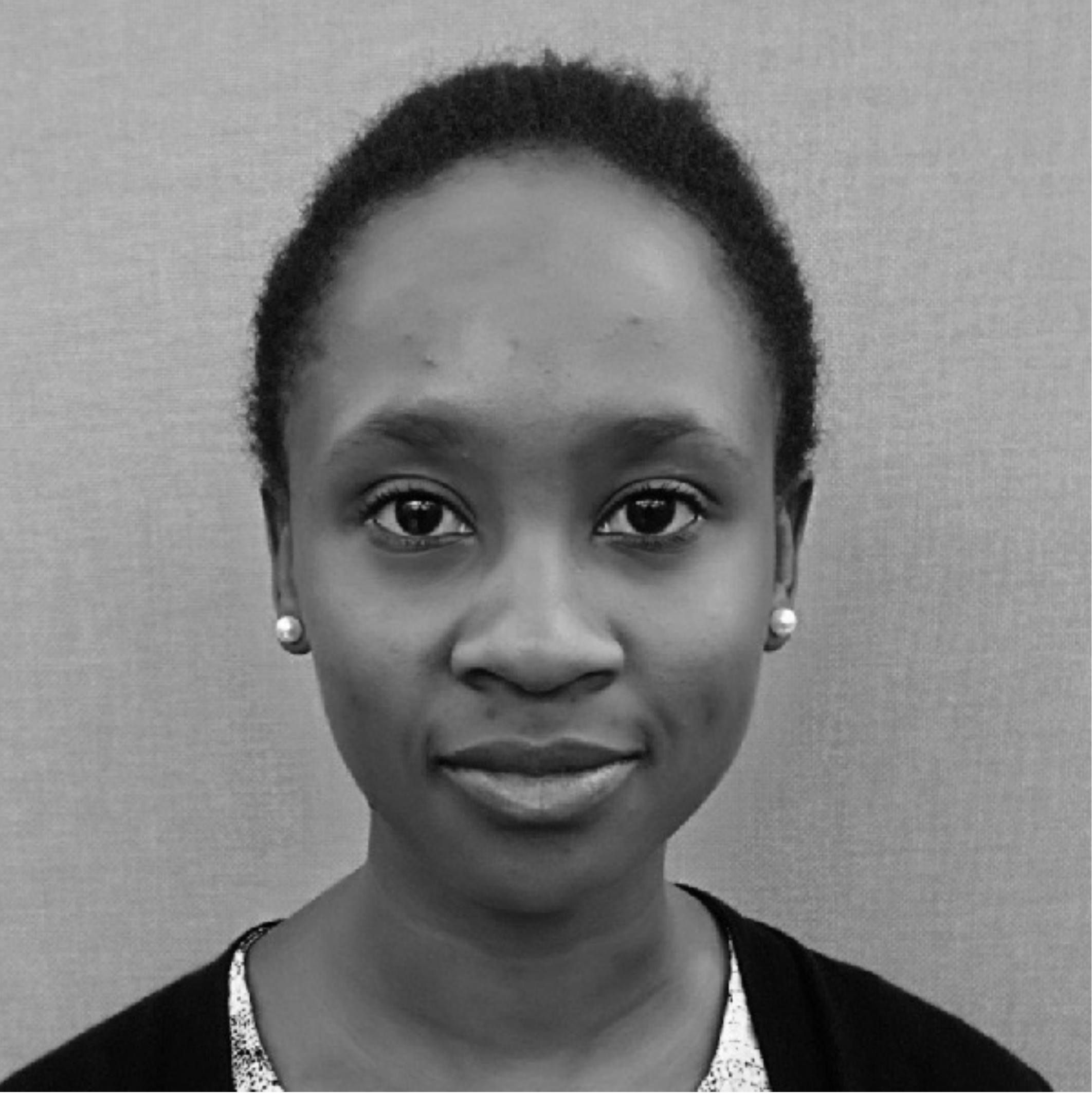}}]{Steffi O. Muhanji} 
Steffi is currently a 5th-year PhD Candidate at the Laboratory for Intelligent Integrated Networks of Engineering Systems (LIINES). Her research interests are in renewable energy integration, transactive energy, the energy-water nexus and distributed control. She has a B.A. in Physics with a Computer Science minor from Vassar College and a B.E. with a focus on energy systems from Thayer School of Engineering. 
\end{IEEEbiography} 
\vspace{-0.7in}

\begin{IEEEbiography}
[{\includegraphics[width=1in,height=1.25in,clip,keepaspectratio]{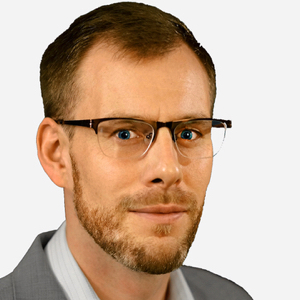}}]{Samuel Golding} 
Samuel V. Golding, President of Community Choice Partners Inc., has been a technical consultant and campaign strategist in the Community Choice Aggregation (CCA) industry for over a decade. He is recognized as a pioneer of the joint action governance structures and advanced operating models that enable CCAs to animate retail markets and develop regulatory frameworks conducive to demand flexibility, particularly in California and New Hampshire. He received his B.A. in International Political Economy in 2007 from The Colorado College.
\end{IEEEbiography}
\vspace{-0.7in}
\begin{IEEEbiography}
[{\includegraphics[width=1in,height=1.25in,clip,keepaspectratio]{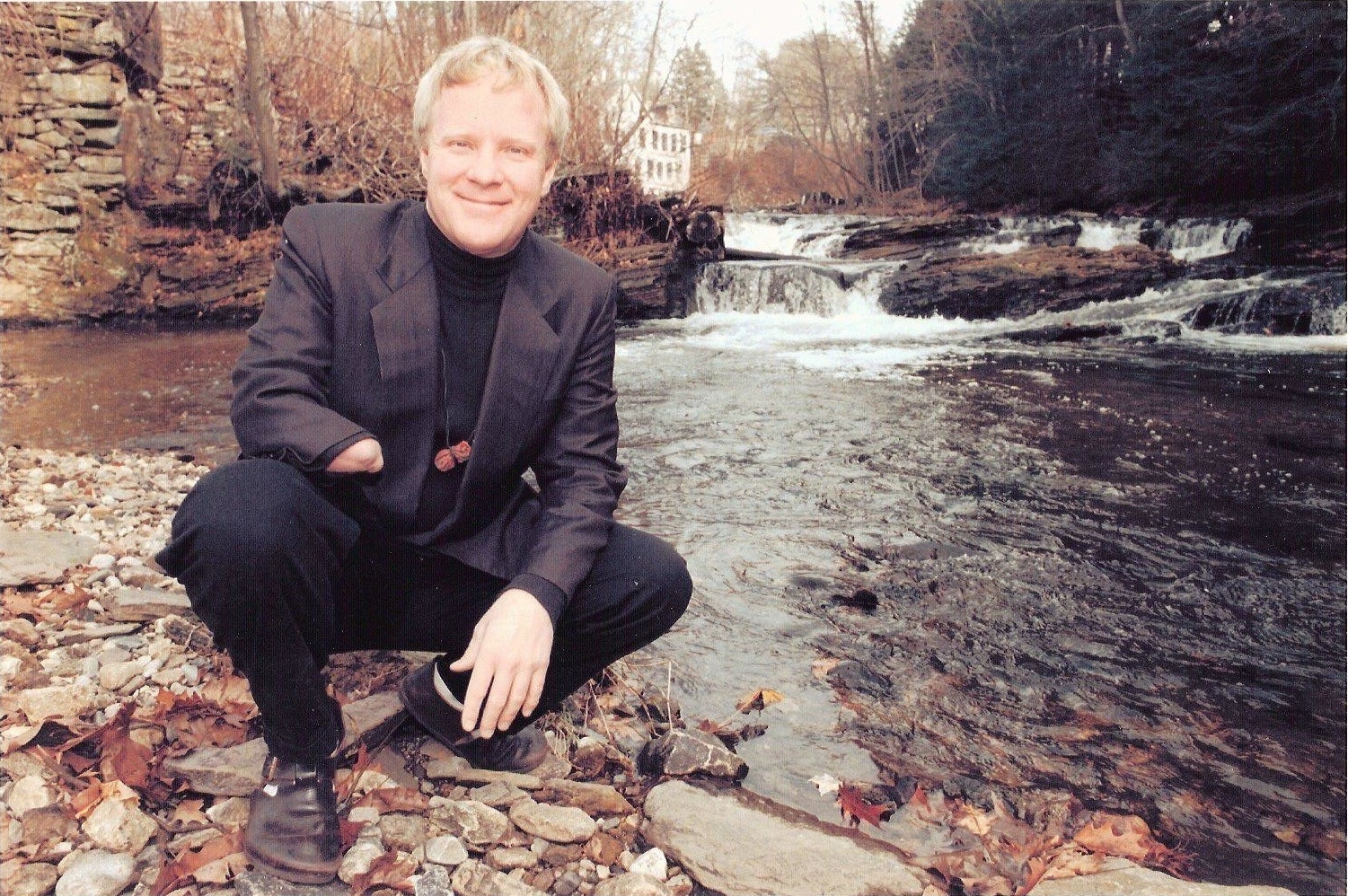}}]{Tad Montgomery} 
Tad Montgomery is the Energy and Facilities Manager for the city of Lebanon, NH.  His responsibilities include assisting the city in meeting its greenhouse gas reduction goals in line with the Paris Climate Accord.  Projects include adoption of ~800 kW of solar power, development of the Lebanon Community Power municipal aggregation program, thermal energy conservation throughout city buildings, and demand reduction in the big electric accounts at the water and wastewater plants. He has an B.S. in Ceramic Engineering from Alfred University and an M.S. in Environmental Systems Analysis (abd) from Humboldt State University.
\end{IEEEbiography}
\vspace{-0.6in}

\begin{IEEEbiography}
[{\includegraphics[width=1in,height=1.1in,clip,keepaspectratio]{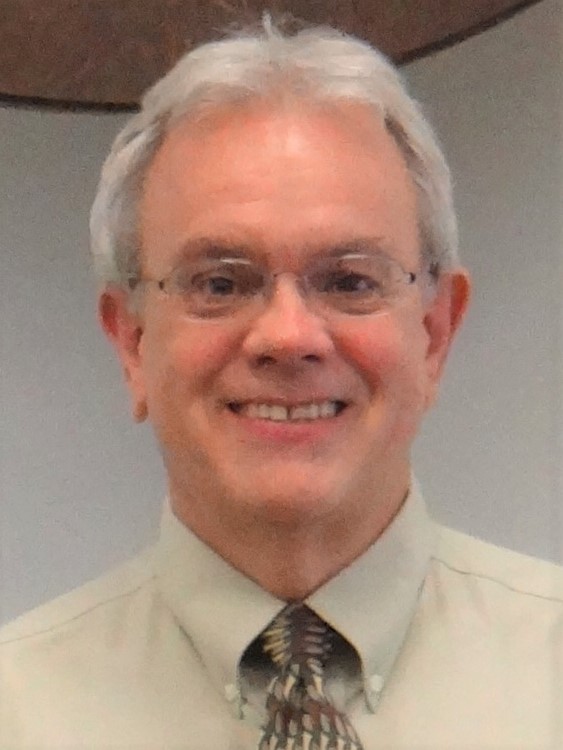}}]{Clifton Below} 
Clifton Below is the Assistant Mayor of the City of Lebanon and Chair of its Energy Advisory Committee. He formerly served 6 years as a Public Utilities Commissioner for NH and 12 years in the NH House and Senate. He earned his B.A. from Dartmouth College in 1980 and a Master of Science in Community Economic Development from Southern NH University in 1985.
\end{IEEEbiography}
 \vspace{-0.7in}
\begin{IEEEbiography}[{\includegraphics[width=1in,height=1.25in,clip,keepaspectratio]{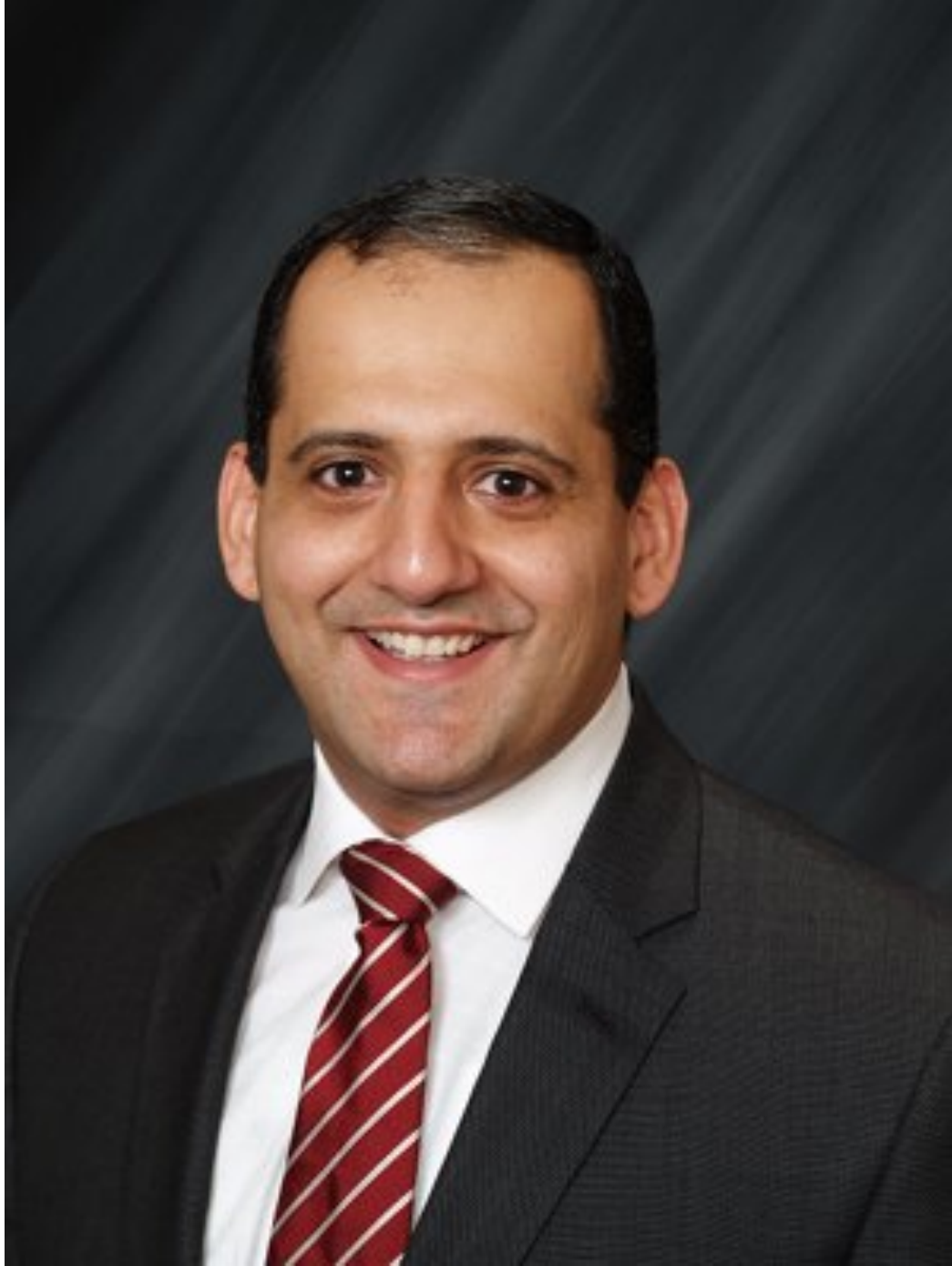}}]{Amro M. Farid} 
Prof. Amro M. Farid is currently an Associate Professor of Engineering at the Thayer School of Engineering at Dartmouth and Adjunct Associate Professor of computer science at the Department of Computer Science. He leads the Laboratory for Intelligent Integrated Networks of Engineering Systems  (LIINES). The laboratory maintains an active research program in Smart Power Grids, Energy-Water Nexus, Energy-Transportation Nexus, Industrial Energy Management,  and Integrated Smart City Infrastructures. He received his Sc. B. in 2000 and his Sc. M. 2002 in mechanical engineering from MIT and his Ph.D. degree in Engineering from the U. of Cambridge (UK).
\end{IEEEbiography}
\end{document}